\documentclass[12pt]{article}
\usepackage[cp1251]{inputenc}
\usepackage[english]{babel}
\usepackage{graphicx,graphics}
 0
\begin{document}

\title{A dressing of zero-range potentials and
electron-molecule scattering problem at low energies}
\author{  S.B. Leble
\small \\ Theoretical Physics and Mathematical Methods Department,
\small\\ Technical University of Gdansk, ul, Narutowicza 11/12, Gdansk,  Poland,
\small \\ leble@mifgate.pg.gda.pl \\  \\[2ex]
S. Yalunin \small\\ Theoretical Physics Department, \small\\
Kaliningrad State University, A. Nevsky st. 14, Kaliningrad,
Russia, \small \\ yalunin@bk.ru}

\maketitle

\renewcommand{\abstractname}{\small Abstract}
\begin{abstract}
A dressing of a nonspherical potential, which includes $n$ zero
range potentials, is considered. The dressing technique is used to
improve ZRP model. Concepts of the partial waves and partial
phases for non-spherical potential are used in order to perform
Darboux transformation. The problem of scattering on the regular
$\hbox{X}_n$ and $\hbox{YX}_n$ structures is studied. The
possibilities of dressed ZRP are illustrated by model calculation
of the low-energy electron-Silane ($\hbox{SiH}_4$) scattering. The
results are discussed.
\center{Key words: multiple scattering,
silane, zero range potential.}
\end{abstract}

\thispagestyle{empty}

\section{Introduction}
The ideas of zero range potential (ZRP) approach were recently
developed to widen limits of the traditional treatment
by Demkov and Ostrovsky $\cite{DO1975}$ and
Albeverio et al. $\cite{AlGa1988}$.
The advantage of the theory is the possibility of obtaining
an exact solution of scattering problem.
The ZRP is conventionally represented as the boundary condition
on the matrix wavefunction at some point.
Alternatively, the ZRP can be represented as pseudopotential
(Breit $\cite{Breit}$).

On the other hand, Darboux transformation (DT) allows to construct
in natural way exactly solvable potentials. General starting point
of the theory goes up to Matveev theorem (see \cite{Mat}). The
transformation can be also defined on the base of covariance
property of the Schr\"{o}dinger equation with respect to a
transformation of wavefunction and potential (Matveev and
Salle $\cite{MatSall}$). Darboux formulas in multi-dimensional
space could be applied in the sense of Andrianov, Borisov and
Ioffe ideas \cite{ABI}. In the circumstances, DT technique can be
used so as to correct ZRP model.

We attempt to dress the ZRP in order to improve the possibilities
of the ZRP model. We use notations and some results from
\cite{LeZa}. DT modifies the generalized ZRP (GZRP) boundary
condition (Section $\ref{S1}$) and creates a potential with
arbitrarily disposed discrete spectrum levels for any angular
momentum $l$. In the section $\ref{S2}$ we consider
$S$-representation for a non-spherical potential so as to dress a
multi-centered potential, which includes $n$ ZRPs. As an important
example, we consider electron scattering by the $\hbox{X}_n$ and
$\hbox{YX}_n$ structures within the framework of the ZRP model
(Section $\ref{S4}$). In section $\ref{S5}$ we present the our
calculations for the electron-$\hbox{SiH}_4$ scattering and
discuss them.

\section{Zero range potential dressing} \label{S1}
Let us start from the simplest case of a central field.
Then angular momentum operator commutates with Hamiltonian and
therefore wavefunction $\Psi({\bf r})$
can be expanded in the spherical waves
\begin{equation}
\Psi({\bf r})=4\pi (kr)^{-1} \sum_{l=0}^{\infty}\sum_{m=-l}^{l}
e^{i(\eta_{l}+l\pi/2)}\ \psi_{l}(r,k)Y_{lm}({\bf n})Y_{lm}^*({\bf n}_0),
\end{equation}
where ${\bf n}=r^{-1}{\bf r}$, ${\bf n}_0$ is initial particle
direction,
$\psi_l(r,k)$ are partial waves, and
$\eta_l$ are phase shifts.
Consider the radial Schr\"{o}dinger equation for partial wave with
angular momentum $l$. The atomic units are used throughout the
present paper, i.e. $\hbar=m=1$ and Born radius
$a_0=\hbar^2/me^2=1$.
\begin{equation} \label{E}
H_l\psi_l(r,k)=
k^2 \psi_l(r,k),
\end{equation}
\begin{equation}\label{EQHl}
H_l=-D^2 +2u_l(r),
\end{equation}
$D=d/dr$ denotes differential operator, and $H_l$ are
Hamiltonian operators of the partial waves.
This equations describe scattering of a particle with energy $E=k^2/2$.
The wavefunctions $\psi_l(r,k)$ at infinity have the form
\begin{equation} \label{BCinf}
\psi_l(r,k)\stackrel{\rm r \rightarrow \infty}{\sim}
\sin\left(kr-\frac{l\pi}{2}+\eta_l(k)\right).
\end{equation}
Let us consider GZRP in coordinate origin.
This potential is conventionally represented as boundary condition
on the wavefunction
(see $\cite{Balt}$)
\begin{equation} \label{Bound}
D^{2l+1}\left(r^{l}\psi_l(r,k)\right)_{r=0}=
-\frac{\alpha_l (2l+1)!}{(2l+1)!!(2l-1)!!}
\left( r^{l}\psi_l(r,k)\right)_{r=0},
\end{equation}
where $\alpha_l$ are inverse
scattering lengths. The potential $u_l=l(l+1)/2r^2$ and therefore
wavefunctions $\psi_l(r,k)$ can be
expressed in terms of the spherical functions
\begin{equation} \label{Eq_psi_h}
\psi_l(r,k)=
\cos(\eta_l)j_l(kr)+\sin(\eta_l)n_l(kr),
\end{equation}
where spherical functions $j_l(x),n_l(x)$ are related to
usual Bessel functions as $j_l(x)=\sqrt{\frac{\pi x}{2}}J_{l+1/2}(x)$,
$n_l(x)=(-1)^l\sqrt{\frac{\pi x}{2}}J_{-l-1/2}(x)$. In
the vicinity of zero they have the asymptotic behavior
$j_l(x)\sim x^{l+1}/(2l+1)!!$, and $n_l(x)\sim x^{-l}(2l-1)!!$.
To substituting the equation
$(\ref{Eq_psi_h})$ into the boundary condition we obtain the
elements of $S$-matrix
\begin{equation}
\exp(2i\eta_l)=\frac{\alpha_l-ik^{2l+1}}{\alpha_l+ik^{2l+1}}.
\end{equation}
The bound states correspond to the poles of the $S$-matrix (i.e
the zeros of the denominator $\alpha_l+ik^{2l+1}$), which lie on
the imaginary positive semi-axis of the complex $k$-plane. It is
obvious that bound state, with orbital momentum $l$, exists only
if $(-1)^l\alpha_l>0$ (elsewise an antibound state exists) and has
the energy $E_l=\alpha_l^{2/(2l+1)}/2$.

Thus, spectral problem for GZRP
is solved for any value $k$.
On the other hand, the equations (\ref{E})
are covariant with respect to DT
that yields the following
transformations of the potentials (coefficients of the operator $H_l$)
\begin{equation} \label{DTpot}
u_l^{(1)}=u_l-Ds_l,
\end{equation}
and the wavefunctions $\psi_l(r,k)$
\begin{equation} \label{DTpsi_l}
\psi_l^{(1)}=\frac{s_l-D}{\sqrt{k^2+b_l^2}}\psi_l(r,k),
\hspace{8mm}
s_l=\frac{D\varphi_l(r,ib_l)}{\varphi_l(r,ib_l)}
\end{equation}
where $\varphi_l(r,ib_l)$ are some
solutions of the equations $(\ref{E})$ at $k=ib_l$, and
$b_l$ are real parameters, which can be both positive or negative.
The DT $(\ref{DTpsi_l})$ combines the solutions
$\psi_l(r,k)$ and a solution $\varphi_l(r,ib_l)$ that corresponds
to another eigen value $k^2=-b_l^2$.
Repeating the
procedure we obtain a chain of the integrable potentials $u_l^{(n)}$.
In general, dressed potential
$u_l^{(1)}$ is real for real function $\varphi_l(r,ib_l)$.

The next step in the dressing procedure of the zero-range
potential ($u_l=l(l+1)/2r^2$) is a definition of the free
parameters of the solutions $\varphi_l(r,ib_l)$. Suppose the prop
functions $\varphi_l(r,ib_l)$ satisfy the boundary conditions
$(\ref{Bound})$ with $\alpha_l=e_l$. In the simplest case of $l=0$
we have
\begin{equation} \label{propf}
\varphi_0(r,ib)=\sinh(br)-(b/e)\cosh(br),
\end{equation}
and
$$
s_0 \stackrel{\rm r \rightarrow \infty}{\sim} -e + (b^2-e^2)r.
$$
The DT $(\ref{DTpsi_l})$
gives rise to the following requirement on dressed wavefunction
$$
D\psi_0^{(1)}= \left(
\frac{b^2+k^2}{\alpha-e}+e\right)
\psi_0^{(1)}.
$$
The dressed potential $u_0^{(1)}$ is given by
$$
u_0^{(1)}=-\frac{b^2(b^2-e^2)}{(b\cosh(br)-
e\sinh(br))^2},
$$
It is regular on semiaxis only if $(b/e) \overline{\in} [0,1)$. In
the limiting case at $b\rightarrow 0$ we obtain long-range
interaction $\propto (1-e r)^{-2}$, which can be regular on
semiaxis only if $e<0$. Assuming $e=\pm b$ we get $u_0^{(1)}=0$
(trivial transformation), and boundary condition can be obtained
by the substitution:
$$
\alpha \rightarrow \frac{b^2+k^2}{\alpha \mp b}\pm b.
$$
To dress free wave $(\alpha=\infty)$
we obtain ZRP at the coordinate origin. Thus, ZRP can be also
introduced in terms of DT.
To consider transformation with parameter
$e=\alpha$ we obtain regular solution $\psi_0^{(1)}$
and tangent of phase shift is
$$
\tan \eta_0^{(1)}=\frac{(e-b)k}{be+k^2}.
$$

In the other cases asymptotic of the functions $s_l$ at zero is
given by
$$
s_l\stackrel{\rm r \rightarrow 0}{\sim}\left\{
\begin{array}{ll}
\displaystyle
-\frac{l}{r}-\frac{b^2_l r}{(2l-1)},
& e_l\neq \infty,\ l>0\\
\displaystyle \frac{l+1}{r}+\frac{b_l^{2^{\mathstrut}} r}{(2l+3)},
& e_l=\infty.
\end{array}
\right.
$$
It is clear that the each DT introduces short-range core of
centrifugal type (which depends on angular momentum $l$) in the
potential. In this situation the boundary conditions on the
dressed wavefunctions $\psi_l[1]$ require modification. Thus, in
the case $e_l\neq \infty,\ l>0$ the boundary conditions become
$$
D^{2l-1}\left( r^{l-1}\psi_l^{(1)}\right)_{r=0}=-
\frac{\alpha_l(2l-1)!}{(k^2-b_l^2)(2l+1)!!(2l-3)!!}
\left( r^{l-1}\psi_l^{(1)}\right)_{r=0}
$$
and in the case $e_l=\infty$ we obtain
$$
D^{2l+3}\left( r^{l+1}\psi_l^{(1)}\right)_{r=0}=-
\frac{\alpha_l(2l+3)!}{(k^2-b_l^2)(2l+3)!!(2l-1)!!}
\left( r^{l+1}\psi_l^{(1)}\right)_{r=0}.
$$

In the generalized case, ZRP with angular momentum $l$ generates
also $2l$ complex poles of the $S$-matrix, which correspond the
quasi-stationary states (resonances).
The DTs $(\ref{DTpsi_l})$ with the parameters $b_l$ results in the
$S$-matrix elements for dressed GZRP
$$
\exp(2i\eta_l^{(1)})=\frac{\alpha_l-ik^{2l+1}}{\alpha_l+ik^{2l+1}}
\left(\frac{b_l-ik}{b_l+ik}\right).
$$
We can use Darboux transformation in order to add (or remove)
poles of the $S$-matrix.

\section{Multi-center problem} \label{S2}
The principal observation allows to built
a zero-range potential eigen function in the multi-center problem.
Let us consider $n$ ZRPs at the points ${\bf r}_i$ and
interaction $v^{(1)}$.
The wavefunction $\Psi({\bf r})$ can be expressed in terms of the
(outgoing-wave) Green function,
defined by the equation
$$
\left(-\nabla^2+2v^{(1)}-k^2\right)
g^{+}({\bf r},{\bf r}')=4\pi
\delta({\bf r}-{\bf r}')
$$
The second (ingoing-wave) Green function is defined by
$g^-({\bf r},{\bf r}')=g^{+*}({\bf r},{\bf r}')$.
The partial waves $\Psi_{\lambda}({\bf r},k)$, defined by
\begin{equation}
\Psi_{\lambda}({\bf r},k)\stackrel{\rm r \rightarrow \infty}{\sim}
\frac{1}{2ikr}\left(
A_{\lambda}({\bf n}) e^{ikr+i\eta_{\lambda}}+
A_{\lambda}(-{\bf n}) e^{-ikr-i\eta_{\lambda}}
\right),
\end{equation}
for multi-centered target which includes $n$ ZRPs and interaction
$v^{(1)}$ can be expressed as a superposition of the Green
functions
\begin{equation} \label{parZRP}
\displaystyle \Psi_{\lambda}({\bf r},k) =
(2ik)^{-1} \sum_{i=1}^{n} c_{\lambda i}
\left( e^{i\eta_{\lambda}} g^+({\bf r},{\bf r}_i) +
e^{-i\eta_{\lambda}} g^-({\bf r},{\bf r}_i) \right)
\end{equation}
in which $\eta_{\lambda}$ are phase shifts,
$A_{\lambda}({\bf n})$ denote $S$-matrix orthonormal
eigenfunctions, $c_{\lambda i}$ are real numbers.
They naturally generalize the spherical partial waves
$(kr)^{-1}\psi_l(r,k) Y_{lm}({\bf n})$ for a non-spherical
potential $\cite{DemRud}$. Expanding the partial waves
$\Psi_{\lambda}({\bf r},k)$ at the infinity we obtain the
expressions for $A_{\lambda}({\bf n})$
\begin{equation}
A_{\lambda}({\bf n})=\sum_{i=1}^{n}
c_{\lambda i}
\Psi^{(1)}({\bf r}_i,-k{\bf n}).
\end{equation}
where $\Psi^{(1)}({\bf r},k{\bf n}) \stackrel{\rm r \rightarrow
\infty}{\sim} e^{ik{\bf n}\cdot {\bf r}}+F^{(1)} e^{ikr}/r$ is
wavefunction for potential $v^{(1)}$, and $F^{(1)}$ is scattering
amplitude. Scattering amplitude for potential $v^{(1)}$ and $n$
ZRP is given by
\begin{equation}
F({\bf n},{\bf n}_0)=F^{(1)}({\bf n},{\bf n}_0)+
\frac{4\pi}{2ik} \sum_{\lambda} \left(
e^{i\eta_{\lambda}}-1\right)
A_{\lambda}({\bf n}) A_{\lambda}^*({\bf n}_0).
\end{equation}
By the imposition of the boundary conditions $(\ref{Bound})$ the
calculation of the partial waves is reduced to the solution of
the following system
\begin{equation} \label{system}
\sum_{j \neq i} c_{\lambda j} \left( s({\bf r}_i,{\bf r}_j) +
c({\bf r}_i,{\bf r}_j) \tan \eta_{\lambda}  \right)= -c_{\lambda
i} \left( k+\delta k_i + (\alpha_i+\delta \alpha_i) \tan
\eta_{\lambda} \right),
\end{equation}
in which we use
\begin{equation}
g^+({\bf r},{\bf r}_i)
\stackrel{{\bf r} \rightarrow {\bf r}_i}{\sim}
\frac{1}{|{\bf r}-{\bf r}_i|}+
\delta \alpha_i + i(k+\delta k_i),
\end{equation}
where $s({\bf r}_i,{\bf r}_j)=\hbox{Im}\, g^+({\bf r}_i,{\bf
r}_j)$, $c({\bf r}_i,{\bf r}_j)=\hbox{Re}\, g^+({\bf r}_i,{\bf
r}_j)$. The tangents $\tan \eta_{\lambda}$ can be found from
compatibility condition of this system. In simplest case $v=\delta
\alpha_i=\delta k_i=0$
$$
c({\bf r}_i,{\bf r}_j)=\frac{\cos(kr_{ij})}{r_{ij}},\hspace{8mm}
s({\bf r}_i,{\bf r}_j)=\frac{\sin(kr_{ij})}{r_{ij}},\hspace{8mm}
r_{ij}=|{\bf r}_i-{\bf r}_j|,
$$
the system $(\ref{system})$ is reduced to the usual equations of
ZRP theory $\cite{DemRud}$. In order to construct functions
$s({\bf r}_i,{\bf r}_j)$ and $c({\bf r}_i,{\bf r}_j)$ for dressed
potential, i.e. for $v^{(1)}$, we need to write down Green
function as single-center expansion over spherical harmonics. In
the simplest case, when the prop function is
$\varphi_0(r,ib)=\sinh(br)$, the Green function is given by
\begin{equation} \label{GF}
g^+({\bf r},{\bf r}')=\frac{e^{ik|{\bf r}-{\bf r}'|}}
{\scriptstyle |{\bf r}-{\bf
r}'|}+\frac{\psi^{(1)}_0(r,k)f_0^{(1)}(r',k)}
{krr'}-\frac{\sin(kr)e^{ikr'}}{krr'},\hspace{7mm}r<r',
\end{equation}
where
$$
\displaystyle f_0^{(1)}(r,k)=\frac{b\coth(br)-ik}{\sqrt{k^2+b^2}}
e^{ikr}, \hspace{8mm} \psi_0^{(1)}(r,k)=\frac{b\coth(br)\sin(kr)-k
\cos(kr)}{\sqrt{k^2+b^2}}.
$$
The function $\psi_0^{(1)}(r,k)$ has the following
asymptotic at infinity
$$
\psi_0^{(1)}(r,k)\stackrel{\rm r \rightarrow \infty}{\sim}
\sin(kr+\delta),\hspace{8mm} \delta=-\arctan(k/|b|).
$$
The integral cross section can be readily derived using
the optical theorem:
$
\sigma=4\pi k^{-1} \hbox{Im} F({\bf n}_0,{\bf n}_0).
$
Thus, averaged integral cross section is given by
\begin{equation}
\bar{\sigma}=\frac{4\pi}{k^2}\sin(\delta)^2+ \frac{4\pi}{k^2}
\sum_{\lambda} \sin(\eta_{\lambda})^2.
\end{equation}

\section{$\hbox{X}_n$ and $\hbox{XY}_n$ structures} \label{S4}
For purpose of illustration we consider scattering
problem for a dressed multi-center potential.
The multi-center scattering within the framework of the ZRP model
was investigated by Demkov and Rudakov $\cite{DemRud}$
(8 centers, cube), Drukarev and Yurova $\cite{DrYur}$
(3 centers in line),
Szmytkowski $\cite{Szmyt}$ (4 centers, regular tetrahedron).

\subsection{Electron$-\hbox{X}_n$ scattering problem}
Let structure $\hbox{X}_n$ contains $n$ identical scatterers,
which involve only $s$ waves. Denote position vector of the
scatterer $j$ by the ${\bf r}_j$. Suppose, for the sake of
simplicity, the distance between any two scatterers $i$ and $j$ is
$r_{ij}\equiv R$. There are three such structures in
three-dimensional space - dome $\hbox{X}_2$, regular trihedron
$\hbox{X}_3$, regular tetrahedron $\hbox{X}_4$. The partial waves
$(\ref{parZRP})$ and phases $\eta_{\lambda}$ can be classified
with respect to symmetry group representation, degeneracy being
defined by the dimension of the representation $\cite{DemRud}$.
The structures $\hbox{X}_2$, $\hbox{X}_3$, $\hbox{X}_4$ belong to
the $D_{\infty h}, C_{3v}, T_d$ point groups respectively. The
equation $(\ref{system})$ leads to algebraical problem
\begin{equation} \label{sys}
\sum_{j=1}^n c_{\lambda j} (\sin(kR) + \tan\eta_{\lambda}\cos(kR)
)= -c_{\lambda i}(kR +   \tan(\eta_{\lambda}) \alpha R) ,
\end{equation}
The phases can be readily found from
compatibility condition of this system.
To factorize the determinant
we derive the expressions for the phases
\begin{equation} \label{phases_Xn}
\tan \eta_1= -\frac{kR+(n-1) \sin (kR)}{\alpha R + (n-1)\cos(kR)},
\end{equation}
\begin{equation}
\tan \eta_{\lambda}= -\frac{kR - \sin(kR)}{\alpha R -
\cos(kR)},\hspace{8mm} \lambda=\overline{2,n}.
\end{equation}
Thus, assuming $n=4$ we obtain the phases of a regular
tetrahedron $\cite{Szmyt}$.
The partial and integral cross sections can be expressed as
\begin{equation}
\sigma_{\lambda}=\frac{4\pi}{k^{2}}
\frac{\tan(\eta_{\lambda})^2}{1+\tan(\eta_{\lambda})^2},\hspace{9mm}
\sigma=\sigma_{1}+(n-1)\sigma_{n}.
\end{equation}

\subsection{Electron$-\hbox{YX}_n$ scattering problem}

The structures $\hbox{YX}_n$ can be used, for instance,
to study a slow electron scattering by the
polyatomic molecules like $\hbox{H}_2\hbox{O}$, $\hbox{NH}_3$,
$\hbox{CH}_4$, etc.
Let ${\bf r}_1,..,{\bf r}_n$ denote the position vectors of
the scatterers $\hbox{X}$ and ${\bf r}_{n+1}$ denotes the
position vector of the scatterer $\hbox{Y}$.
The scatterers $\hbox{X}$ are situated in
vertices of a regular structure $\hbox{X}_n$, i.e.
$r_{ij}\equiv R$.
Suppose, for the sake of simplicity, the distance between
the scatterer $\hbox{Y}$ and any scatterer $\hbox{X}$ is
$r_{i(n+1)}\equiv D$. Therefore the
position of the scatterer $\hbox{Y}$ perfectly fixed only if $n=4$
(geometric center of the tetrahedron, $R=2\sqrt{\frac{2}{3}}D$).
The partial waves are given by the equation $(\ref{parZRP})$, where
the summation should be performed over the $i=\overline{1,n+1}$.
The constants $c_{\lambda i}$ and phases can be derived analytically.
Thus, we obtain the phases
\begin{equation}
\tan \eta_{\lambda}=-\frac{kR-\sin(kR)}{\alpha R - \cos(kR)}, \ \
\ \ \ \lambda=\overline{3,n+1},
\end{equation}
where $\alpha$ is inverse scattering length of
the scatterers $\hbox{X}$.
The $x_{1,2}=\tan \eta_{1,2}$ obey the quadratic equation
\begin{equation}
(k+\beta x) \left(x+\frac{kR+(n-1)\sin (kR) }{\alpha R +(n-1)\cos
(kR)}\right)= n\frac{R(\sin (kD) + x \cos (kD))^2}{D^2(\alpha R +
(n-1)\cos (kR))},
\end{equation}
where $\beta$ indicates inverse scattering length of the scatterer
$\hbox{Y}$. In the limiting case when the distance between
$\hbox{X}$ and $\hbox{Y}$ scatterers is very large,
i.e. $D\rightarrow \infty$, the expression for $\tan\eta_{1}$
becomes $(\ref{phases_Xn})$
and $\tan\eta_{2}\rightarrow -k/\beta$.
This situation corresponds to
independent scattering on the structure $\hbox{X}_n$ and scatterer
$\hbox{Y}$. The tangent of $\eta_1$ also reduces to $\tan \eta_1$ for
structure $\hbox{X}_n$ in the limit $\alpha_0\rightarrow \infty$.

\subsection{Dressing of the structures $\hbox{X}_n$ and
discussion}  \label{S5}
\begin{figure}
\centering
\includegraphics[scale=0.9]{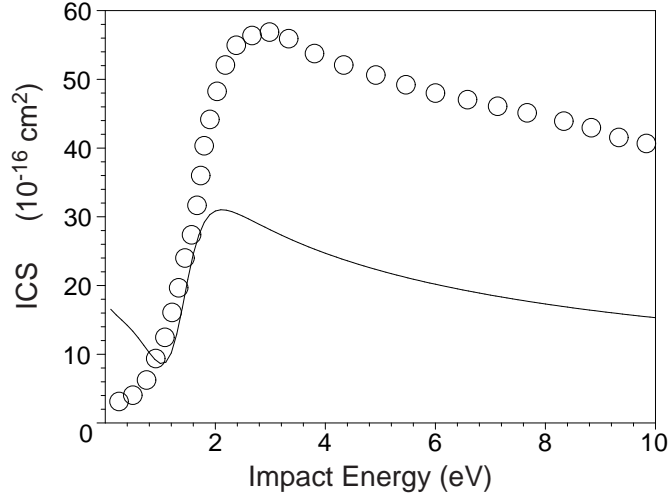}
\parbox[t]{0.55\textwidth}{
\caption{Integral cross sections for electron$-\hbox{SiH}_4$
scattering. ZRP model: solid line;
Experimental results of Wan et al.$\cite{Wan}$: open circles.}
\label{f:ics_SiH4}}
\end{figure}

\begin{figure}
\centering
\includegraphics[scale=0.9]{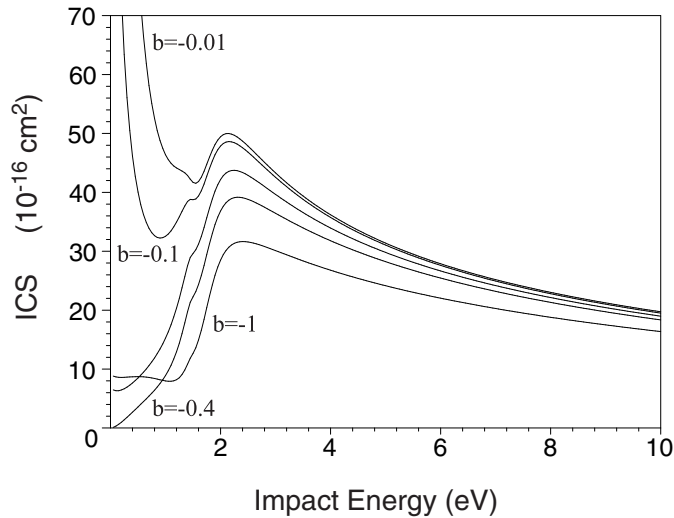}
\parbox[t]{0.6\textwidth}{
\caption{Integral cross sections for electron$-\hbox{SiH}_4$
scattering on dressed ZRPs}
\label{f:icsd_SiH4}}
\end{figure}

The integral cross sections for $\hbox{SiH}_4$
(closed-shell ground-state $^1A_1$)
are plotted in fig. $\ref{f:ics_SiH4}$
for a number of values of $\alpha, \beta,R,D$, which are regarded
as constant in the range of interest.
The present calculations were carried out with
$\alpha=0.33\, a_0^{-1},\ \beta=0.41\,
a_0^{-1},\ R=4.51\, a_0,\ D=2.76\, a_0$.
Our calculations were made within the framework of the ZRP model and
hence they are not expected to be correct for low impact energies
(i.e for energies $\sim 1$ eV) where polarization effects
($\hbox{SiH}_4$ spherical polarizability $\sim 30.4\, a_0^3$)
are known to be important. Because induced polarization potentials
are always attractive, including  polarization algebraically
increases the computed phases ($\eta_{1,3}<0$ at $E \sim 0$)
and therefore decreases the integral cross section.
Thus, fig. $\ref{f:ics_SiH4}$ clearly shows
that polarization effects reproduce the deep minimum
(Ramsauer-Townsend minimum near $0.35$ eV) seen in
the experimental data $\cite{Wan}$ and numerical calculations
$\cite{Jain}$, which are incomparably smaller
in the our calculations.
The ICS can be corrected by the ZRPs dressing.
Fig. $\ref{f:icsd_SiH4}$ show ICS for dressed $\hbox{YX}_4$ structure.
In our calculation, the parameters $\alpha=0.35$, $\beta=0.38$
were used. In the higher collision energies our
cross section (fig. $\ref{f:ics_SiH4}$)
differs from the experimental results in size but
coincides in shape.
It is known that role of the higher partial waves is important
at the higher collision energies.
Thus, taking into consideration the $p$ waves for the scatterers
$\hbox{X}$ and $\hbox{Y}$ we can add in partial waves and
improve the agreement both in size
at the higher energies and in the position of shape resonance.

\section{Conclusion}
We demonstrate the posibilities of DT in multi-center scattering
problem. Thus, these transformations allow to correct the ZRP model
at low energies. In the limiting case $b\rightarrow 0$ DT
induces the long-range forces $\sim r^{-3}$ depending on
angular momentum and leads to singularity in the cross section
at zero energy.

\end{document}